\documentclass[fleqn,10pt]{wlscirep}
\usepackage[utf8]{inputenc}
\usepackage[linesnumbered,ruled,vlined]{algorithm2e} 
\usepackage[T1]{fontenc}
\usepackage{graphicx}
\usepackage{amsmath} 
\usepackage{mathtools} 
\usepackage{subfig}
\usepackage{tabularx}
\usepackage{multirow,siunitx}
\usepackage{adjustbox}
\usepackage{caption}
\usepackage{array}
\newcolumntype{P}[1]{>{\centering\arraybackslash}p{#1}}
\newcolumntype{M}[1]{>{\centering\arraybackslash}m{#1}}
\title{PathoNet: Deep learning assisted evaluation of Ki-67 and tumor infiltrating lymphocytes (TILs) as prognostic factors in breast cancer; A large dataset and baseline}

\author[1,7]{Farzin Negahbani}
\author[1]{Rasool Sabzi}
\author[2]{Bita Pakniyat Jahromi}
\author[3]{Fatemeh Movahedi}
\author[2]{Mahsa Kohandel Shirazi}
\author[3]{Shayan Majidi}
\author[4]{Dena Firouzabadi}
\author[5,6,*]{Amirreza Dehghanian}

\affil[1]{Department of Computer Science and Engineering, Shiraz University, Shiraz, Iran.}
\affil[2]{Department of Pathology, Shiraz University of medical Science, Shiraz, Iran.}
\affil[3]{Shiraz University of medical Science, Shiraz, Iran.}
\affil[4]{Department of Clinical Pharmacy, Shiraz University of Medical Sciences, Shiraz, Iran}
\affil[5]{Molecular Pathology and Cytogenetics division, Department of Pathology, Shiraz University of Medical Sciences, Shiraz, Iran.}
\affil[6]{Trauma Research Center, Shiraz University of Medical Sciences, Shiraz, Iran.} 
\affil[7]{Department of Computer Science and Engineering, Koc University, Istanbul, Turkey.}
\affil[*]{adehghan@sums.ac.ir}

\begin{abstract}
The nuclear protein Ki-67 and Tumor infiltrating lymphocytes (TILs) have been introduced as prognostic factors in predicting tumor progression and its treatment response. The value of Ki-67 index and TILs in approach to heterogeneous tumors such as Breast cancer (BC), known as the most common cancer in women worldwide, has been highlighted in literature. Due to the indeterminable and subjective nature of Ki-67 as well as TILs scoring, automated methods using machine learning, specifically approaches based on deep learning, have attracted attention. Yet, deep learning methods need considerable annotated data. In the absence of publicly available benchmarks for BC Ki-67 stained cell detection and further annotated classification of cells, we propose SHIDC-BC-Ki-67 as a dataset for aforementioned purpose. We also introduce a novel pipeline and a backend, namely PathoNet for Ki-67 immunostained cell detection and classification and simultaneous determination of intratumoral TILs score. Further, we show that despite facing challenges, our proposed backend, PathoNet, outperforms the state of the art methods proposed to date in the harmonic mean measure. Dataset is publicly available in \url{http://shidc.ir} and all experiment codes published in \url{https://github.com/SHIDCenter/PathoNet}.
\end{abstract}
\begin{document}
\flushbottom
\maketitle
\thispagestyle{empty}

\section*{Introduction}

The nuclear protein Ki-67 was first detected in Hodgkin lymphoma cell line and introduced as an active protein in proliferating cells rather than quiescent ones.\cite{gerdes1983production} It was further confirmed that the monoclonal antibody Ki-67 is present during all cell cycle phases except for the G0.\cite{gerdes1984cell} However variation in its intensity has been reported during different phases, as for the G1 having the lowest Ki-67 expression.\cite{lopez1991modalities}Knowing that excessive cellular proliferation correlates with malignancy, this protein marker can be beneficial in identifying high-grade tumors and can also convey prognostic value in approach to tumor management. Breast cancer, the most common cancer in women worldwide, is a heterogeneous type of cancer.\cite{siegel2015cancer} The heterogeneity of breast cancer makes it difficult in terms of approach to its therapy and markers such as Ki-67 have been introduced as prognostic factors in predicting progression of this type of tumor.\cite{dowsett2008emerging,jones2009prognostic,taneja2010classical} The established method for Ki-67 detection is Immunohistochemical (IHC) analysis using MIB-1 as the most commonly used antibody in the staining process on fresh or paraffin embedded tissue.\cite{urruticoechea2005proliferation,dowsett2011assessment} Considering that the marker's final estimation is based on the percentage of positively stained cells amongst all tumoral cells counted by an expert pathologist, this makes its assessment process somewhat subjective. On the other hand, to increase the accuracy of estimation, it has been suggested to count all tumor cells in dynamic samples. If impossible to do so, at least 500-1000 cells in the representative areas of the whole section are recommended to be counted by the pathologist.\cite{dowsett2011assessment} This, in turn, can be very time consuming for large numbers of samples facing variation due to its subjective bias. Another limitation in the assessment of Ki-67 has been the absence of a single cut-off point for indicating its expression. Taking into account that host immunity can play a role in the progression and treatment of tumor, tumor infiltrating lymphocytes (TILs) has also been proposed as a predictive factor in terms of response to therapy in breast cancer.\cite{denkert2010tumor,denkert2018tumour,mao2014value,mao2016prognostic} The absence of unanimous cut-off points for both Ki-67 and TILs and the probability of bias due to their subjective nature urges the need for an exact continuous calculation of both factors, which motivated us to tackle this problem using Artificial intelligence (AI).

\begin{figure}[ht]
\centering
\includegraphics[width=\linewidth]{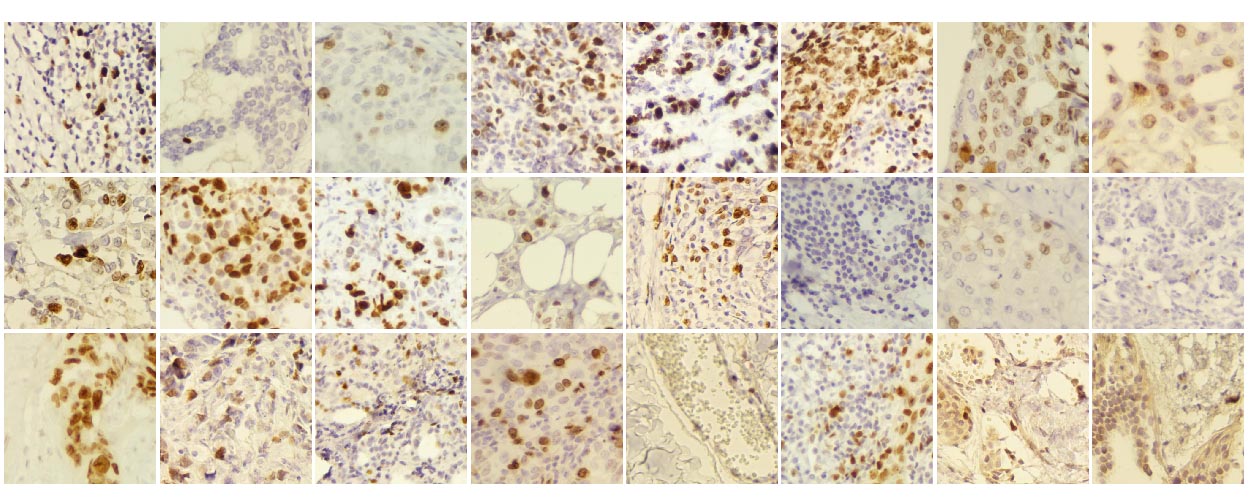}
\caption{SHIDC-B-Ki-67 dataset samples.}
\label{fig:shidc_samples}
\end{figure}

AI had a significant impact on improving speed and precision in clinical diagnosis and is becoming an inseparable part of the medical community with vast applications.\cite{kononenko1997application } Before introducing deep learning, conventional AI algorithms were commonly used; however, designing a generalized and robust method needs field experts to extract handcrafted features. By the advent of deep networks, having the facility of automatically learning the best features from the input data, this issue has almost been resolved. In addition, if these algorithms are developed and trained with diverse and adequate data, they are generalizable and robust. Convolutional neural network (CNN) is a class of deep networks widely used in different areas such as Robotics, Bioinformatics, Computer Vision, etc. and showed significant performance, especially in image processing \cite{soanssa,haskins2020deep,hafiz2020survey}. 

Lecun et al. first used CNNs \cite{doi:10.1162/neco.1989.1.4.541} for handwritten digit classification in 1989, though, for years, due to computation barriers and lack of sufficient data, CNNs haven't been used. By the appearance of GPUs in 2012, computation capabilities have been immensely improved, and enough data provided in ImageNet competition\cite{ILSVRC15} lead to the advent of AlexNet\cite{krizhevsky2012imagenet}, which won the competition using a CNN approach. To overcome the shortcomings of manual assessment of Ki-67 and yet to take advantage of the probable beneficial role of Ki-67 and TILs in approach to breast cancer, in this experimental study we have suggested the use of artificial intelligence assisted methods with emphasis on deep learning in accurate detection of tumoral cells and  Ki-67 scoring  and TILs expression in breast cancer samples. 

The intention of performing this study and its different aspects can be summarized into four categories. First, we introduce a dataset with detection and classification annotation in three different classes that provide a benchmark for Ki-67 stained cell detection, classification, and proliferation index estimation. Second, we suggest a novel pipeline that can achieve cell detection and classification and further evaluate the proposed pipeline on our benchmark. Third, we recommend a deep network, namely PathoNet, that outperforms the state of the art backends with the proposed pipeline in ki-67 immunopositive, immunonegative, and lymphocyte detection and classification. Lastly, we introduce a residual inception module that provides higher accuracy without causing vanishing gradient or overfitting issues.

\section*{Literature Review}
Data regarding cell proliferation estimation based on Ki-67 detection utilizing deep learning and conventional machine learning algorithms are present in the literature. Some conventional methods have been suggested in this regard; A study on neuroendocrine tumors (NET) presented a framework for Ki-67 assessment of NET samples that can differentiate tumoral from non-tumoral cells (such as lymphocytes) and furthermore classified immunopositive and immunonegative tumor cells to achieve automatic Ki-67 scoring.\cite{xing2013automatic} Lastly, we investigate related datasets that are publicly available. For the tumor biopsies of meningiomas and oligodendrogliomas based on immunohistochemical (IHC) Ki-67 stained images, Swiderska et al., introduced a combination of morphological methods, texture analysis, classification, and thresholding.\cite{swiderska2015hot} Shi et al. carried out a study based on morphological methods to address color distribution inconsistency of different cell types in the IHC Ki-67 staining of nasopharyngeal carcinoma images. They suggested classifying image pixels using local pixel correlations taken from specific color spaces.\cite{shi2016automated} Geread et al. proposed a robust unsupervised method for discriminating between brown and blue colors. \cite{geread2019ihc}

Despite improvements, conventional methods not only lack generalization and accuracy compared to direct interpretation by pathologists they are also complex and cumbersome to develop because of having handcrafted features. As for Deep Learning methods, different approaches tackling image classification, cell detection, nuclei detection, and Ki-67 proliferation index estimation in histopathological images have been reported. Xu et al. suggested using deep learning features in multiple instances learning (MIL) framework for colon cancer classification.\cite{xu2014deep} Weidi et al. proposed a cell spatial density map for an image using convolutional neural networks (CNNs) for the task of automated cell counting and detection in the existence of cell clumping or overlapping\cite{weidi2015microscopy}.

On the other hand, Cohen et al. suggested redundant counting instead of estimating a density map. Moreover, they introduced a network stemmed from inception networks called Count-ception for cell counting.\cite{ paul2017count} Spanhol et al. evaluated deep features for the breast cancer recognition task to further compare to conventional methods.\cite{spanhol2017deep} In another study on Ki-67 scoring by Saha et al., they automatically estimate hotspots using a gamma mixture model, using a deep learning model with a decision layer. Their method assesses the Ki-67 proliferation index in breast cancer immunohistochemical images.\cite{saha2017jiang} Zhang et al. used convolutional neural networks (CNN) to classify images as benign or malignant and a single shot multibox detector as an object detector to assess Ki-67 proliferation score in breast biopsies of patients with breast cancer.\cite{zhang2018tumor} Sornapudi et al. extracted localized features by taking advantage of superpixels generated using clustering algorithms. Afterward, they applied a CNN to perform nuclei detection on extracted features.\cite{sornapudi2018deep} Due to the restriction of manually labeled Ki-67 datasets, Jiang et al. proposed a new model consisting of residual modules and Squeeze-and-Excitation block named small SE-ResNet, which has fewer parameters in order to prevent the model from over-fitting. Similar classification accuracy was reported for SE-ResNet compared to the ResNet in classifying samples into benign and malignant.\cite{jiang2019breast} Liu et al. tackled cell counting problem as a regression problem by producing cell density map in a preprocessing step and further utilized a stacked deep convolutional neural network model for counting.\cite{liu2019novel} 

Available datasets in form of publicly presented benchmarks can be divided into the benign-malignant classification and cell counting categories. For benign-malignant image classification, Spanhol et al. introduced BreakHis, which consists of breast cancer histopathological images obtained from partial mastectomy specimens from 82 patients with four different magnifications. \cite{spanhol2016breast} Diverse cell counting and nuclei detection datasets such as synthetically generated VGG-CellS \cite{lempitsky2010learning}, real samples of human bone marrow by Kainz et al.\cite{kainz2015you}, Modified Bone Marrow (MBM) and human subcutaneous adipose tissue (ADI) datasets by Cohen et al.\cite{paul2017count}, and Dublin Cell Counting (DCC) proposed by Marsden et al.\cite{marsden2018people} are all of the many examples of datasets presented. However, none of the mentioned benchmarks provide facilities for both cell detection and classification. To the best of our knowledge, SHIDC-B-Ki-67 is the first benchmark introducing  immunohistochemically marked BC specimens that has cell annotations in three different classes of immunopositive, immuno negative, and tumor infiltrating lymphocytes. 

\section*{Dataset}
The critical role of databank in deep learning is evident to experts in this field, therefore in the absence of a comprehensive Ki-67 marked dataset, SHIDC-B-Ki-67 was gathered by the use of numerous and various data labeled by pathology experts. This dataset contains microscopic tru-cut biopsy images of malignant breast tumors exclusively of the invasive ductal carcinoma type. Images were taken from biopsy specimens gathered during a clinical study from 2017 to 2020. All patients who participated in this study were patients with a clinical diagnosis of breast cancer whose breast tru-cut biopsies were taken at Shiraz University of Medical Sciences’ affiliated hospitals’ pathology Laboratory in Shiraz, Iran. Shiraz University of Medical Sciences institutional review and ethical board committee approved the study (ethics approval ID: IR.SUMS.REC.1399.756)  and written informed consent was gathered from all patients willing to take part in the study. Moreover, all the data were anonymized.

\begin{table}[h]
\centering
\label{tab:my-table}
\begin{tabular}{c|c|c|c|c|c|c|}
\cline{2-7}
\textbf{} &
  \multicolumn{2}{c|}{\textbf{Total set (2357 Img)}} &
  \multicolumn{2}{c|}{\textbf{Training set (1656 Img)}} &
  \multicolumn{2}{c|}{\textbf{Test set (701 Img)}} \\ \hline
\multicolumn{1}{|c|}{\textbf{Cell Type}} &
  \# Cells &
  avg./Img &
  \# Cells &
  avg./Img &
  \# Cells &
  avg./image \\ \hline
\multicolumn{1}{|c|}{Immunopositive} & 50861  & 21.58 & 35106  & 21.19 & 15755 & 22.50 \\ \hline
\multicolumn{1}{|c|}{Immunonegative} & 107647 & 45.69 & 75008  & 45.29 & 32639 & 46.62 \\ \hline
\multicolumn{1}{|c|}{Lymphocyte}     & 4490   & 1.90  & 3112   & 1.87  & 1378  & 1.96  \\ \hline
\multicolumn{1}{|c|}{Total Cells} & 162998 & 23.06 & 113226 & 22.79 & 49772 & 23.70 \\ \hline
\end{tabular}
\caption{Statistics of the annotated cells.}
\end{table}

Images were taken from breast mass Tru-cut biopsy sample slides, which were further stained for Ki-67 by immunohistochemistry (IHC) method. Specific monoclonal antibodies were obtained from Biocare Medical, Ca, USA. The adjuvant detection kit, named Master polymer plus detection system (peroxidase), was obtained from Master Diagnostica, Granada. Dimethylbenzene (Xylene 99.5\%) obtained from Samchun Chemical Co., Ltd, South Korea, Ethanol 100\% from JATA Co., Iran, and 96\% from Kimia Alcohol Zanjan Co., Iran, EDTA and Tris (molecular biology grade) obtained from Pars Tous Biotechnology, Iran. Phosphate buffer saline (PBS 1X) 0.01M, pH 7.4 was prepared. At first, paraffin-embedded blocks of breast tissue were sectioned (4-5 microns) and fixed on glass slides. Prepared slides were further used for the immunostaining procedure. Hematoxylin color was used for counterstaining to perform nuclear staining and semi-quantify the extent of immunostaining that will further be evaluated. Accordingly, the expert pathologists identified the tumoral areas in each slide, by visual analysis of tissue sections under a light microscope. The final diagnosis of each case was also approved by two experienced pathologists and confirmed by some ancillary tests such as immune staining for more markers. An Olympus BX-51 system microscope with a relay lens with a magnification of 10 Œ coupled to OMAX microscope digital color camera A35180U3 was used to get digital images of the tumoral tissue slides. Complementary details of the camera and setup are provided in the supplementary information document. Images acquired in RGB  (24-bit color depth, 8 bits per color channel) color space using magnifying factor 400x, corresponding to objective lens 40x. The stepwise acquisition of images at different magnifications is as follows: first, the pathologist identifies the tumor and defines a region of interest (ROI). In order to cover the whole ROI, several images are captured. The pathologist preferentially selects images of the tumoral area, but some of the images also include transitional parts, e.g., tumoral/non-tumoral areas. A final visual (i.e., manual) inspection discards out-of-focus images. Stained images were labeled by expert pathologists as Ki-67 positive tumor cells, Ki-67 negative tumor cells, and tumor cells with positive infiltrating lymphocytes separately. Fig.\ref{fig:shidc_samples} depicts some samples of SHIDC-B-Ki-67 dataset.

\begin{figure}[ht]
  \centering
  \subfloat[Raw image.]{\includegraphics[width=0.2\textwidth]{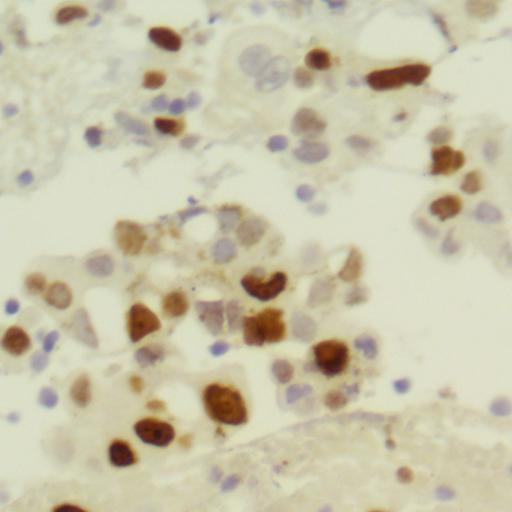}\label{fig:labeling_raw_img}}
  \hfill
  \subfloat[Expert annotations on SHIDC-Lab]{\includegraphics[width=0.3\textwidth]{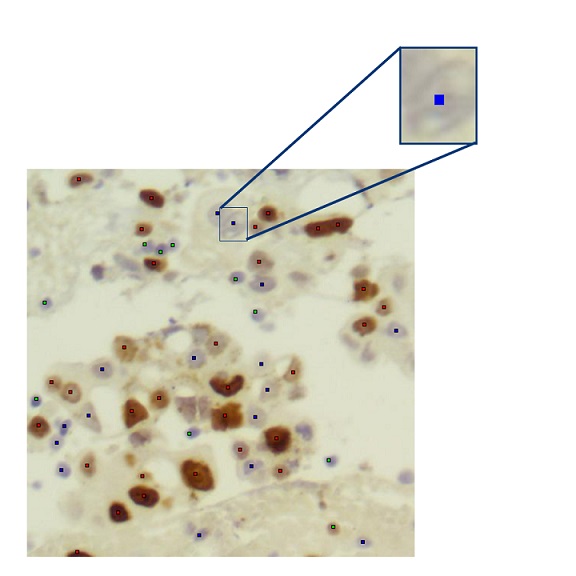}\label{fig:labeling_labeled_img}}
  \hfill
  \subfloat[Extracted labels.]{\includegraphics[width=0.3\textwidth]{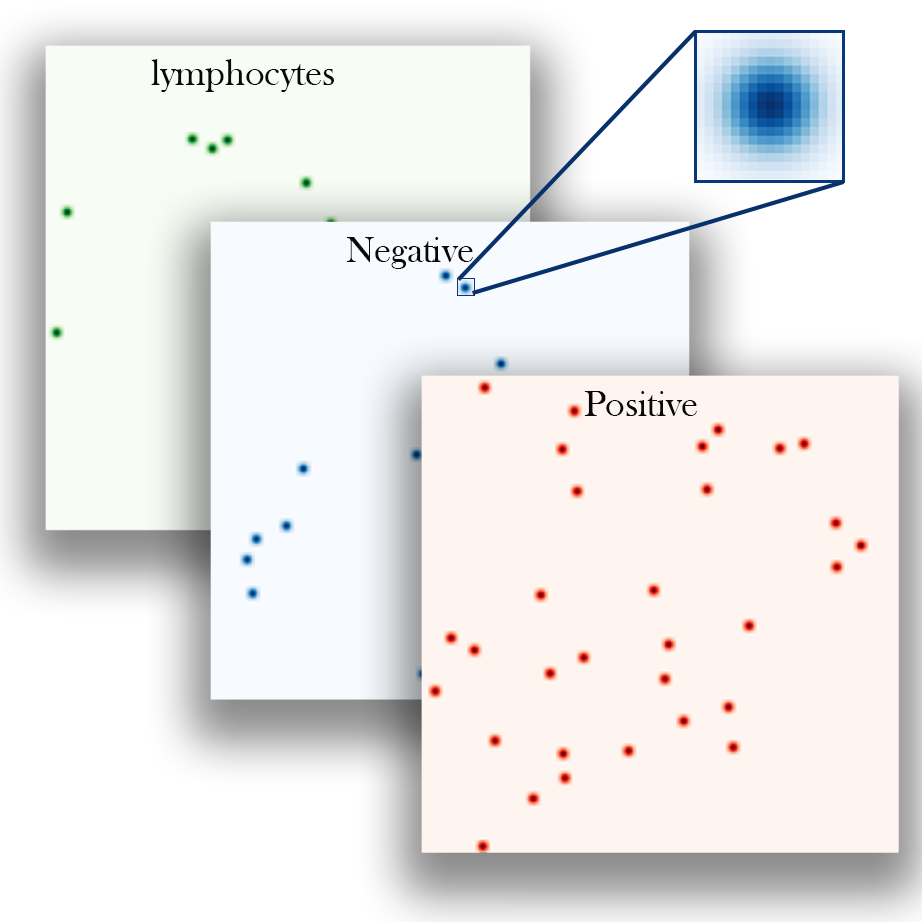}\label{fig:labeling_labels}}
  \caption{Labeling process of SHIDC-B-Ki-67 dataset. First, capture and crop (a)raw image, then (b) experts specify cell centers along with cell types. Lastly, (c) density maps are generated from cell centers.}
\vspace{0.5cm}
\label{fig:labeling}
\end{figure}

\textbf{Labeling.} Precision and quality of expert labels play a vital role in the correct learning process and, respectively, method accuracy; however, labeling real world data is a challenging and labor-intensive task. In SHIDC-B-Ki-67, each image contains 69 cells on average and in total 162998 cells. Manually labeling all cells, wastes the valuable and limited time of experts that should be dedicated to the treatment and assessment of patients. Therefore, SHIDC-Lab, a web-based framework, is designed and implemented. Another big challenge is choosing a label type. In the segmentation task, labeling means to determine a class for each pixel; however, due to overlapping pixels between many cells and since it is infeasible to annotate each pixel on histopathological images scale, this approach is not applicable in our case. In addition, annotations of detection tasks are usually a bounding box around the object of interest. Utilizing this type of annotation in our case where cells are small and abundant with different sizes, makes the network design procedure more complicated. To overcome this issue, cell center plus cell type is picked as the annotation. Fig.\ref{fig:labeling} demonstrates labels in this study.

Although this approach of labeling increases the speed of evaluating marker expression, it is not without limitations. Since just one pixel is picked as the center of each cell, many pixels exist without a label. This makes the data unbalanced and,  in turn, a more laborious learning process is needed. Furthermore, this labeling approach is not appropriate for most of the ordinary neural network loss functions because they cannot be a suitable representative of loss in the task. To clarify, we bring an example in which a network predicts the center of a cell with a 2-pixel or 200-pixel drift compared to the center picked by experts. Normal loss functions cannot discriminate between these two pixels and score them equally. Also, experts’ annotations are error-prone that may cause the same problem. This issue can be addressed by considering an uncertainty for center pixels annotated by the experts. The uncertainty is modeled as a Gaussian distribution with the labeled pixel as the center and n-pixel variance for each cell. As a result, instead of having a 3-channel pixel as the label, a density map for each class is used. Consequently, the nature of the problem is converted into a density map estimation problem.

\section*{Methodology}
In the following sections, we explain our suggested pipeline for cell classification and detection of Ki-67 marked images. The pipeline takes advantage of CNN to extract features and estimate density maps from an input RGB image.

\textbf{UNET.} UNet is one of the most commonly used architectures in biomedical image segmentation\cite{ronneberger2015u} that consists of symmetric U-shaped architecture with two paths. The first path is a standard convolutional neural network named contraction path or encoder fed by input images, and the second one is an expansive path or a so-called decoder. In each layer of the decoder, an up-sampling layer increases the feature map dimension until it gets to the input image size. UNET is a fully convolutional model made from 19 layers. The novelty of this method is in using skip connections between corresponding encoder and decoder layers, transmitting high detail features from encoder layers to the same size layers of the decoder. This approach leads to achieving accurate location results. Similar to many studies motivated by UNet design\cite{myronenko20183d, dolz2018ivd}, we suggest PathoNet as a backend for the proposed pipeline based on UNet architecture.

\textbf{Residual Dilated Inception Module.} Cell detection, classification, and counting in histopathological images is a troublesome task due to the nature of tissues with a variety of cell types and the high possibility of overlapping cells. Nevertheless, it is crucial in the accurate diagnosis of disease and a physicians’ approach to its management. This explains the need for accuracy in developing such networks, which can mostly be done by designing a deeply structured network involving plenty of parameters, yet this mostly causes vanishing gradient issues. 
\begin{figure}[!tbp]
  \centering
  \subfloat[dilation rate=1]{\includegraphics[width=0.25\textwidth]{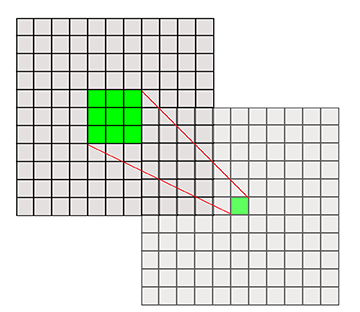}\label{fig:d1}}
  \hfill
  \subfloat[dilation rate=2]{\includegraphics[width=0.25\textwidth]{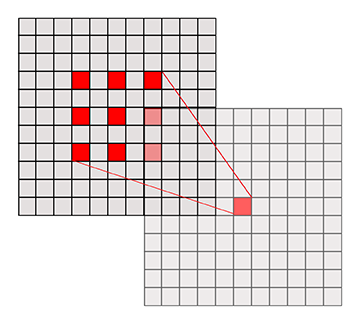}\label{fig:d2}}
  \hfill
  \subfloat[dilation rate=4]{\includegraphics[width=0.25\textwidth]{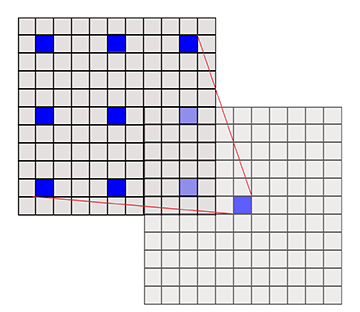}\label{fig:d4}}
  \caption{Dilated convolution kernels with dilation rate (a)1,(b)2, and (c)4.}
  \label{fig:diltedConv}
\end{figure}

On the other hand, cell size may vary from image to image, and since the same cell types usually sit together, picking a suitable kernel size is crucial. Szegedy et al. \cite{szegedy2015going} proposed an inception module to provide a wider field of view without having exponential parameter growth. In the inception model, instead of building a deeper network by stacking convolutional layers, parallel convolutional layers were added to make the network wider. Therefore, by increasing the number of kernels in a layer, higher accuracy can be achieved without facing the vanishing gradient problem. Also, by utilizing different kernel sizes in one module, the problem of choosing a fixed kernel size was solved, and field of view increased. In the inception module proposed by Szegedy et al., three parallel kernels with 5x5, 3x3, and 1x1 sizes were used before using a max-pooling layer. Also, to prevent immense computation growth, input tensor channels were decreased using 1x1 kernels before 5x5 and 3x3 kernels. Still, their method increases network parameters, which means the network needs more data to train and is more likely to be over-fitted. To overcome this issue, Yang et al.\cite{ yang2019dilated} employed dilated convolutions instead of regular convolution layers inside the inception module. D-Dilated convolution has a D distance between each kernel element that can cover a wider region. Fig.\ref{fig:diltedConv} shows this operator with different dilation rates. Eq.\ref{eq:dilation} shows a KxK dilated convolution with step size D.

\begin{equation}
 \hspace{0.3\linewidth} output[i] = \sum_{n=1}^K Input[i+d.n].Kernel[n]   
 \label{eq:dilation}
\end{equation}

\begin{figure}[ht]
  \subfloat[Conventional inception module]{
	\begin{minipage}[c]{
	   0.3\textwidth}
	   \centering
	   \includegraphics[width=1.2\textwidth]{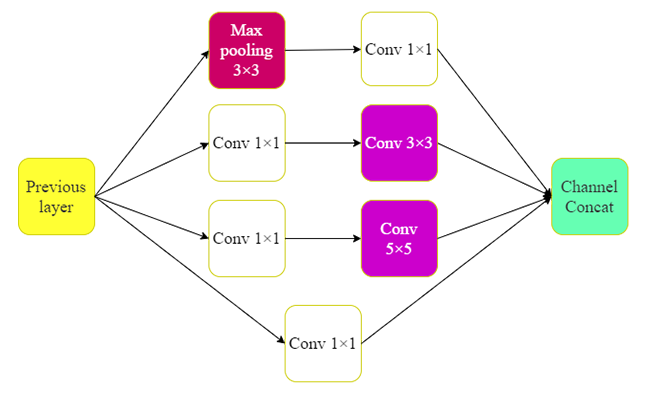}
	\end{minipage}}
 \hfill 
  \subfloat[Residual Dilated Inception Module(Encoder path)]{
	\begin{minipage}[c]{
	   0.35\textwidth}
	   \centering
	   \includegraphics[width=0.5\textwidth]{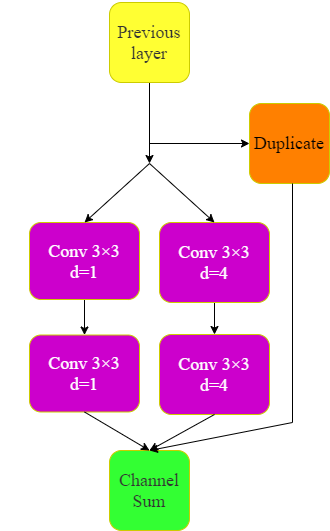}
	\end{minipage}}
  \subfloat[Residual Dilated Inception Module(Decoder path)]{
	\begin{minipage}[c]{
	   0.35\textwidth}
	   \centering
	   \includegraphics[width=0.5\textwidth]{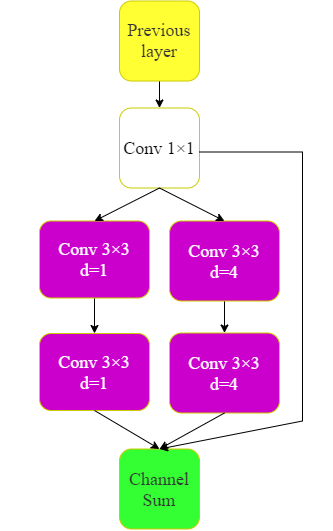}
	\end{minipage}}
\caption{Comparison of conventional and proposed inception module (a)Conventional inception module, (b)Residual Dilated Inception Module(Encoder path), (c)Residual Dilated Inception Module(Decoder path)}
\label{fig:inception}
\end{figure}

In other words, a convolution operation is a dilated convolution with a dilation rate of one. By using dilated convolution in the inception module, Yang et al. maintained the accuracy and reduced the number of model parameters. Also, instead of using multiple 1x1 kernels before other kernels, they just used a 1x1 kernel that shares output with the next kernels. In the parallel convolution, outputs add together while in the inception module proposed by Szegedy et al., these outputs concat together, which consumes more memory. In  \cite{he2016deep}, the authors used ResNet blocks to design deeper architectures without facing overfitting or vanishing gradient effect. In the ResNet blocks, the activation function output of layers sums up with the previous layer’s output. Motivated by the mentioned studies, in this article, we used a new inception module called residual dilated inception module (RDIM). The RDIM consists of two parallel paths where the first path has two convolution layers with kernel size 3x3, and the second one is built by stacking two 3x3 dilated convolution layers with dilation rates equal to 4. In the end, the two paths’ output sums up with the module input. However, since the number of the input channels must be the same with the output to be able to perform summation, two inception modules are used in the encoder and the decoder. In the encoder part, where the input channels are half of the output, duplicated input sums with the result of the two paths. In contrast, in the decoder, a 1x1 convolution with the same kernel size as the two paths applies to the input then sums with the results of parallel routes. Fig.\ref{fig:inception} shows inception and residual dilated inception module structures.

\begin{figure}[ht]
\centering
\includegraphics[width=\linewidth]{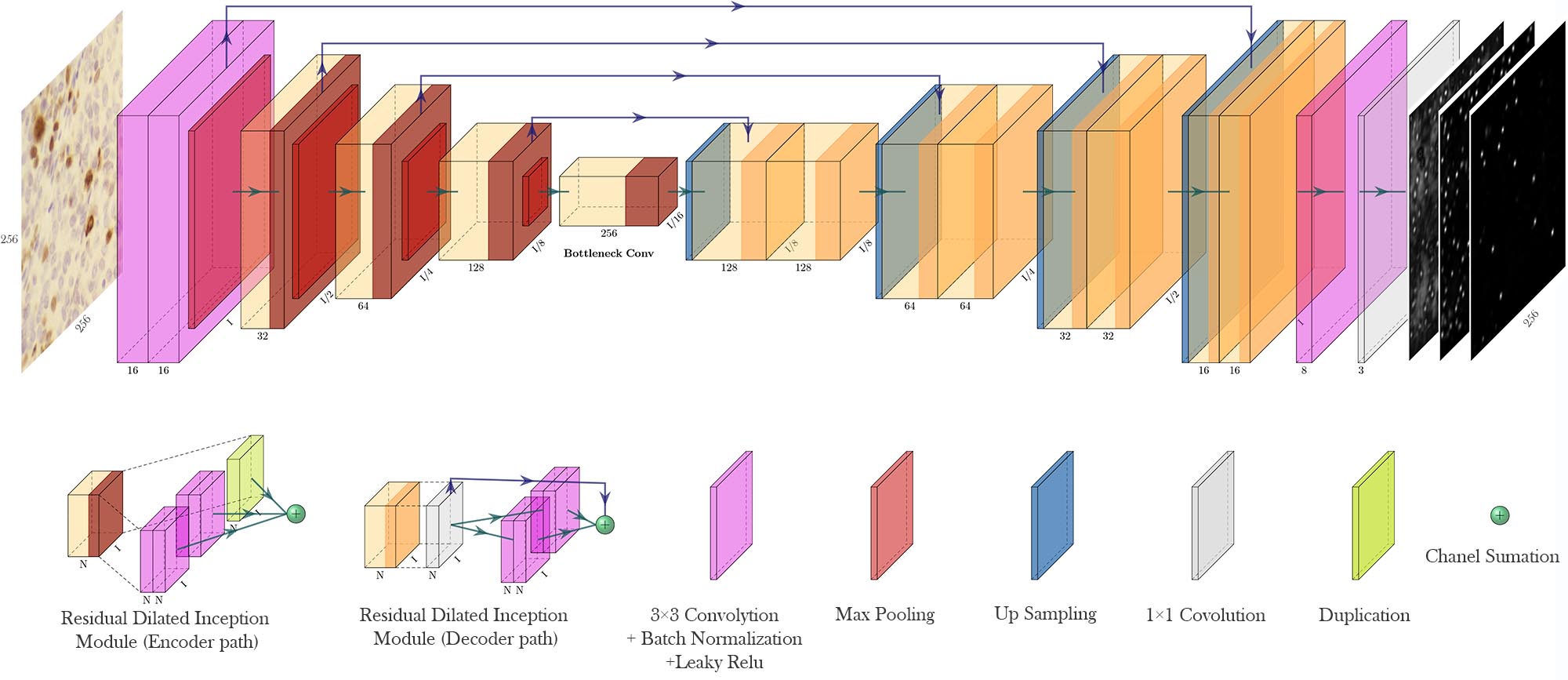}
\caption{PathoNet Architecture.}
\label{fig:pathonet}
\end{figure}

\textbf{PathoNet.} PathoNet first extracts features from input images then predicts candidate pixels for ki-67 immunopositive and immunonegative cells, and also lymphocytes with their corresponding density values. The proposed backend utilizes the UNet-like backbone, where except for the first layer, convolutional layers are replaced by RDIM. In PathoNet, first, input passes through two convolutional layers. Then in the encoder, three and the decoder four RDIMs are used. In the end, a layer consists of three 1x1 convolution layers, and a linear activation function produces a three-channel output of the model. Fig.\ref{fig:pathonet} demonstrates PathoNet architecture. This network results in a three-channel, 256 by 256 matrix that each channel corresponds to the density map of ki-67 immunopositive, immunonegative, or lymphocyte class.

\begin{figure*}[!t]
\begin{adjustbox}{width=0.8\columnwidth,center}
  \begin{minipage}[htp]{0.4\textwidth}
    \centering
        \begin{algorithm}[H]
        \small
        \SetAlgoLined
         Start with all pixels with the lowest possible value \tcp*{Forms the basis for initial watersheds}\
         \For {each intensity level K}{
             \For {each group of pixels of intensity K}{
                 \If{adjacent to exactly one existing}{
                    Add these pixels to that region\;
                 }
                 \ElseIf{adjacent to more than one existing regions}{
                    Mark as boundary\;
                 }
                 \Else{
                    Start a new region\;
                 }
             } 
         }
         \caption{Watershed Algorithm}
        \end{algorithm}
        
  \end{minipage}
  \begin{minipage}[htp]{0.5\textwidth}
    \centering
        \includegraphics[width=\textwidth]{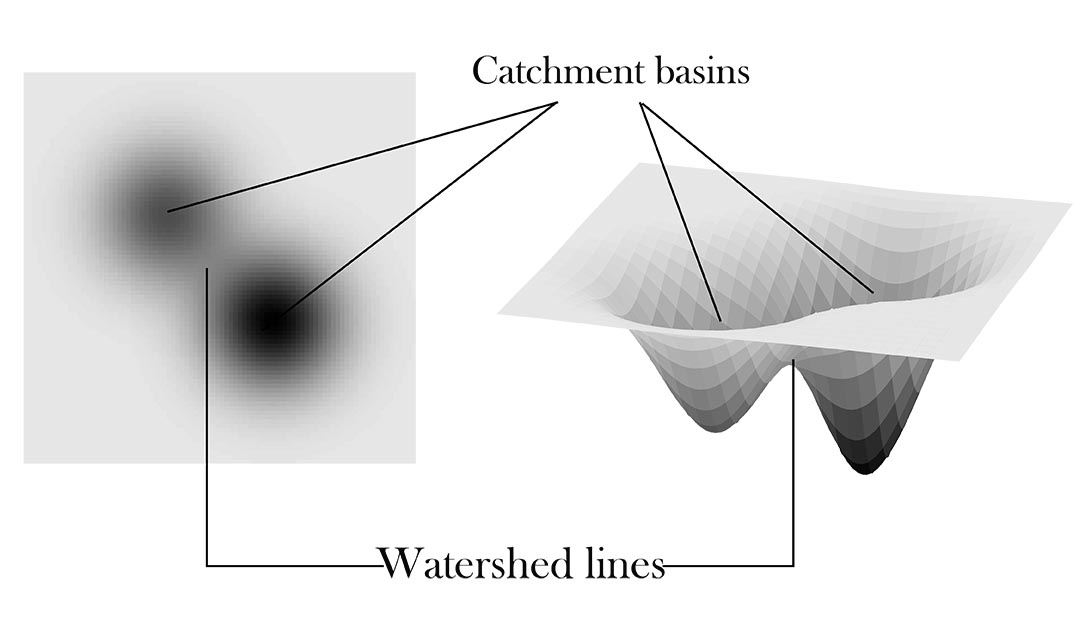}
    \caption{Watershed Algorithm maps grayscale images to a topographic relief space.}\label{fig:watershed_vis}
  \end{minipage}
  \end{adjustbox}
\end{figure*}

\textbf{Watershed.} Watershed algorithm is a conventional method that is useful in medical and material science image segmentation.\cite{atta20163d} Watershed was first introduced in 1978,\cite{Lantujoul1978LaSE } but over the past decades, different versions have been proposed. \cite{ kornilov2018overview }  This algorithm maps grayscale images to a topographic relief space. In a 3-dimensional relief space, each point corresponds to a pixel in the input image with height value equal to the pixel intensity. This relief space consists of different regions, namely low-lying valleys (minimums), high-altitude ridges (watershed lines), and slopes (catchment basins). These regions are demonstrated in Fig.\ref{fig:watershed_vis}. Aim of the Watershed algorithm is to find catchment basins or watershed lines. Watershed is based on a simple yet useful theory. Lines connecting these points are called watershed lines. Watershed lines clarify the segment borders, and the holes are catchment basins or image segments. Algorithm 1 describes a simple Watershed algorithm.

\begin{figure}[thpb]
\centering
\includegraphics[width=\linewidth]{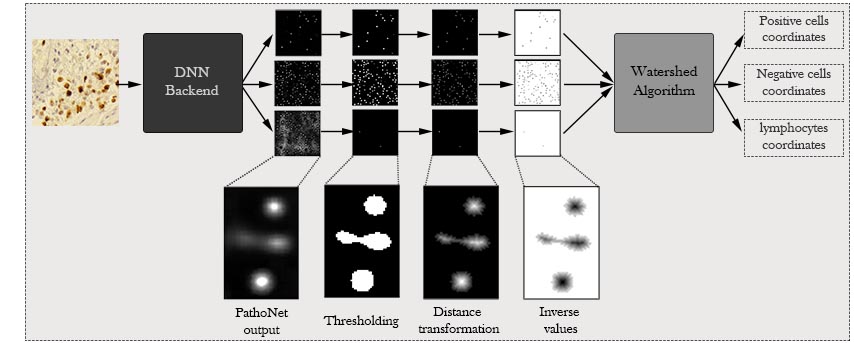}
\caption{Proposed method pipeline. First, a backend, which is a density map estimator based on CNNs, predicts a density map for each class. Then thresholding applies to density maps and produces binary images. Further, inversed distance transformation scores region centers with low and borders with high values. Finally, the watershed algorithm predicts cell centers, and the pipeline outputs the cell coordinates.}
\label{fig:pipeline}
\end{figure}

\textbf{Proposed Pipeline. } The proposed pipeline consists of three components: 1)PathoNet network, 2)post-processing, and 3) Watershed algorithm. The Watershed algorithm and post-processing components do not contain trainable elements, so in the training phase, we train the PathoNet. During the test phase, PathoNet generates a 3-channel density map from an input image where each channel corresponds to the density map of a class. Since there can be multiple pixels in a small region of the map with close or equal densities, choosing a single pixel as the center is rather ambiguous. Besides, due to the existence of noise and low-density points, the amount of false-positive predictions will increase. Hence, a post-processing stage is added to the pipeline. Within the post-processing stage, at first, points with less than the specified threshold are removed and points more than the threshold maps to 255. Second, distance transformation is applied on each channel producing a grayscale image in which the value of points in continuous regions shows the distance from region borders. After applying distance transformation, areas with only one maximum point are produced. For non-circular regions or where we have overlapping cells, there might be multiple maximum points. Finally, we apply a watershed algorithm that produces cell center coordinates to segment these regions and the overlapping cells. It is notable that because Watershed finds minimum values, an inverse operation is applied before performing the watershed method in the third step of the post-processing phase. Fig.\ref{fig:pipeline} presents the proposed pipeline.

\section*{Discussion and Results}

To the best of our knowledge, most of the detection methods that lead to automatic assessment of Ki-67 have reported their results on datasets that are not publicly available, and this field lacks a common ground due to its various samples making it to perform a fair evaluation. Our method also generates pixel coordinates plus class types, which differs from previous studies in this field, which only predicts cell nuclei or just the Ki-67 score.\cite{jiang2019breast,zhang2018tumor} Since to the best of our knowledge, no method exists that can be applied directly on our classification and detection benchmark, to compare and evaluate, we used state of the art methods as the backend in our pipeline and reported the results. DeepLabV3 \cite{chen2017rethinking} that has the best results on the VOC-PASCAL 2012 \cite{Everingham10} was used. The last layer of DeepLabV3 was replaced with a 3-channel convolution layer that has a linear activation function and can have outputs similar to PathoNet. The deep Lab method has Mobilenet and Xeption implementations, which we provided results for both. Similar to PathoNet, FCRN-A, and FCRN-B \cite{ doi:10.1080/21681163.2016.1149104} was designed for density estimation but at a single class. For evaluating FCRN-A and FCRN-B methods as our pipeline backend, their last layer changed from one kernel to three kernels. Since the PathoNet backbone is similar to UNet, we trained it in the same setting as the other methods did, yet, UNet proved to be underfitting and, therefore, could not be evaluated. Evaluation results of the proposed pipeline with different backends is provided in Table \ref{table:performance} .

\textbf{Experimenal Setup.} In this study, Keras\cite{chollet2015keras} framework used to train the network using two NVIDIA Geforce GTX 1060  and an Intel Core-i5 6400. To train PathoNet, MSE loss function by means of the ADAM optimizer is used. The learning rate is set to 0.0001 and decreases with a 0.1 rate every ten epochs. 

\textbf{Measurements.} To evaluate our pipeline, first, we need to define True and False predictions. We count an estimation as True positive (TP) when the predicted center has a less than R pixel distance with the corresponding ground truth; otherwise, it is marked as a False positive(FP). If more than one detected center is within an R-pixel distance with the same cell type in the ground truth, estimation with lower distance counts as TP and otherwise as FP. Finally, cells are defined in the ground truth, but without any prediction for False Negatives (FN). With the given definitions, the precision and recall formulas are shown in Eq.\ref{eq:prec_rec}.
\begin{equation}
 \hspace{0.25\linewidth} Precision = \frac{TP}{TP+FP} \hspace{0.1\linewidth} Recall = \frac{TP}{TP+FN} 
 \label{eq:prec_rec}
\end{equation}  
A model with high recall and low precision rates detects most of the pixels as cell centers, while most of them are FP. On the other hand, low recall and high precision rates in a model bring about the detection of few cells, while most detected cells are TP. None of the mentioned cases is useful; therefore, our goal is to develop a model that holds a trade-off between precision and recall. F1 score or harmonic mean is an appropriate measure that can be used for this evaluation that can be calculated using Eq.\ref{eq:f1_score}.

\begin{equation}
 \hspace{0.3\linewidth} F1-score = 2 . \frac{Precision . Recall}{ Precision + Recall}
  \label{eq:f1_score}
\end{equation}  

In practice, Ki-67 and TILs scores are metrics that can assist oncologists in their approach to patient treatment and help primarily estimate the tumor prognosis. Therefore we also evaluated the introduced pipeline with different backends by means of TILs and Ki-67 scores based on experts’ annotations. To this end, after estimation of TIL and Ki-67 scores for images, performance of models evaluated using RMSE between prediction and annotation scores.

\begin{equation}
 \hspace{0.23\linewidth} 
 Ki-67-Score=\frac{Immunopositive}{Immunopositive+Immunonegative}
  \label{eq:KI_score}
\end{equation}

\begin{equation}
 \hspace{0.18\linewidth}
 TIL-Score = \frac{Lymphocyte}{Lymphocyte + Immunopositive+Immunonegative} 
 \label{eq:TIL_score}
\end{equation}

Second, since each raw image is cropped into smaller ones, we grouped all images that belonged to one patient into one group and classified the estimation into different cut-off categories that have been presented in previous studies. This will be evaluated using classification accuracy. As suggested by Saha et al,\cite{saha2017jiang}, cases with Ki-67 scores below 16 percent are accounted as less proliferative, between 16 and 30 as having average proliferation rate, and higher than 30 as highly proliferative. Also, based on TILs score, the cut-off ranges presented in literature is between 0 and 10, between 11 and 39, and higher than 40 percent\cite{denkert2018tumour}. Table \ref{table:scores} compare RMSE and accuracy of different backends based on the proposed pipeline for Ki-67 and TIL scoring. 

\textbf{Quantitative Results.} As shown in Table \ref{table:performance}, DeepLabv3-Xeption performed better in Ki-67 immunopositive cell detection. However, our introduced model outperforms the others in the detection of Ki-67 immunonegative cells in terms of precision and harmonic mean (F1 Score). The suggested pipeline using FCRN-B has better precision and harmonic mean in the lymphocyte class. In contrast, our model has a better recall rate, meaning that pipeline using PathoNet could detect more labeled lymphocytes than the other methods. The proposed backend performed better in terms of precision by having the lowest FP and outperforming the others in terms of overall F1. Although DeepLabv3-Xception is very close to ours in the overall F1, as shown in Table \ref{table:timing_results}, it has 12 times more parameters compared to the proposed method. So, not only DeepLabv3-Xception needs more computation resources for training, but also, the proposed method can be processed faster and provides higher FPS while maintaining a better F1 score than DeepLabv3-Xception.

\begin{table}[h]
\begin{adjustbox}{width=\columnwidth,center}
\centering
\begin{tabular}{|c|c|c|c|c|c|c|c|c|c|c|c|c|}
\hline
\multirow{2}{*}{\textbf{Model}} & \multicolumn{3}{c|}{\textbf{Immunopositive}} & \multicolumn{3}{c|}{\textbf{Immunonegative}} & \multicolumn{3}{c|}{\textbf{Lymphocyte}} & \multicolumn{3}{c|}{\textbf{Average}}\\ \cline{2-13} 
                           & Prec.\ & Rec.\       & F1\         & Prec.\ & Rec.\ & F1\ & Prec.\      & Rec.\       & F1\ & Prec.\      & Rec.\       & F1\        \\ \hline
Modified DeepLabv3-Mobilenetv2 \cite{chen2017rethinking}  & 0.833 & 0.823 & 0.828 & 0.712 & 0.756 & 0.733 & 0.117 & 0.330 & 0.172 & 0.733 & 0.765 & 0.747  \\ \hline
Modified DeepLabv3-Xeption \cite{chen2017rethinking}      & \textbf{0.838} & 0.862 & \textbf{0.850} & 0.739 & \textbf{0.802} & 0.769 & 0.389 & 0.424 & 0.406 & 0.760 & 0.810 & 0.784  \\ \hline
Modified FCRN-A  \cite{doi:10.1080/21681163.2016.1149104} & 0.812 & \textbf{0.869}  & 0.840 & 0.702 & 0.798 & 0.747 & 0.341 & 0.453 & 0.389 & 0.726 & 0.810 & 0.766  \\ \hline
Modified FCRN-B \cite{doi:10.1080/21681163.2016.1149104}  & 0.823 & 0.868 & 0.845 & 0.727 & 0.797 & 0.760 & \textbf{0.406} & 0.436 & \textbf{0.421} & 0.748 & 0.809 & 0.777  \\ \hline
Proposed Method                                           & 0.819 & 0.866 & 0.842 & \textbf{0.770} & 0.779 & \textbf{0.774} & 0.309 & \textbf{0.541} & 0.394 & \textbf{0.772} & 0.799 & \textbf{0.785}  \\ \hline
\end{tabular}
\end{adjustbox}
\caption{Performance measures for Ki-67 and TIL classification and detection. }
\label{table:performance}
\end{table}

\begin{table}[h]
\centering
\label{tab:scores-table}
\centering
\begin{tabular}{|c|M{2cm}|M{2cm}|M{2cm}|M{2cm}|}
\hline
\textbf{Model} & \textbf{KI-67 score (RMSE)} & \textbf{TIL score (RMSE)} & \textbf{KI-67 cut-off accuracy} & \textbf{TIL cut-off accuracy}\\ \hline
Modified DeepLabv3-Mobilenetv2 \cite{chen2017rethinking}  & 0.050  & 0.086 & 0.956 & 0.739\\ \hline
Modified DeepLabv3-Xeption \cite{chen2017rethinking}      & 0.063 & 0.017  & 0.956 & 0.956 \\ \hline
Modified FCRN-A  \cite{doi:10.1080/21681163.2016.1149104} & 0.067  & 0.020 & 0.956 & 1 \\ \hline
Modified FCRN-B\cite{doi:10.1080/21681163.2016.1149104}  & 0.069  & 0.016  & 0.956 & 1 \\ \hline
Proposed Method                                          & 0.062  & 0.026 & 0.956 & 0.956\\ \hline
\end{tabular}
\caption{TIL and Ki-67 scores measures plus RMSE measures between predicted scores and ground truth score.}
\label{table:scores}
\end{table}

\begin{table}[h]
\centering
\label{tab:run-time-table}
\begin{tabular}{|c|c|c|c|}
\hline
\multirow{2}{*}{\textbf{Model}} & \multirow{2}{*}{\textbf{\# Parameters}} & \multirow{2}{*}{\textbf{FPS}} & \multirow{2}{*}{\textbf{avg. F1}} \\
                                                                           &          &       &       \\ \hline
Modified DeepLabv3-Mobilenetv2 \cite{chen2017rethinking}  & 3236907  & 14.95 & 0.747 \\ \hline
Modified DeepLabv3-Xeption \cite{chen2017rethinking}      & 41253587 & 9     & 0.784 \\ \hline
Modified FCRN-A  \cite{doi:10.1080/21681163.2016.1149104} & 2142019  & 20.87 & 0.766 \\ \hline
Modified FCRN-B\cite{doi:10.1080/21681163.2016.1149104}  & 1365888  & 18.5  & 0.777 \\ \hline
Ours (PathoNet)                                                            & 3142208  & 14.36 & 0.785 \\ \hline
\end{tabular}
\caption{Run-time and number of parameters comparison between different methods.}
\label{table:timing_results}
\end{table}

\begin{figure}[h]
\centering
\includegraphics[width=0.85\linewidth]{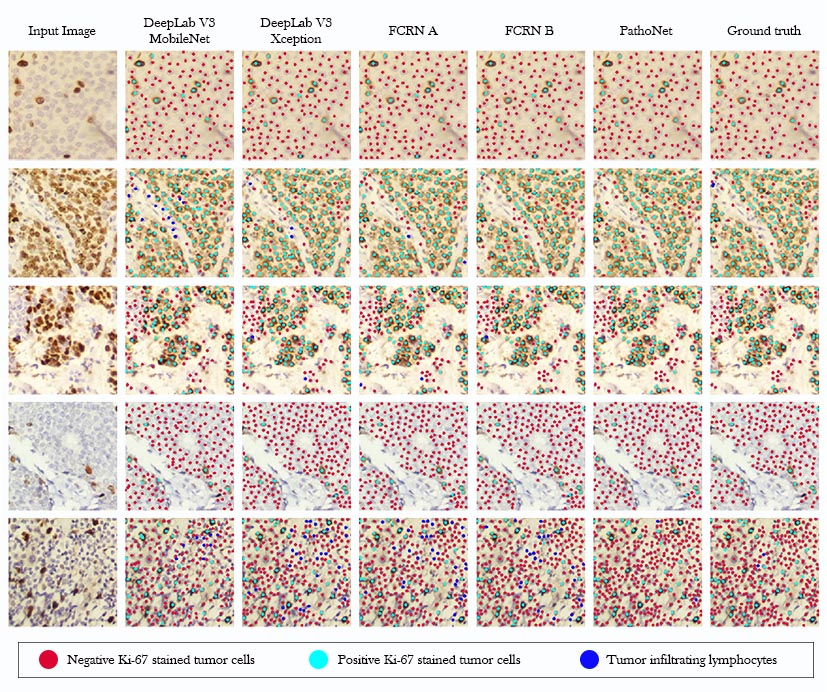}
\caption{Qualitative results of samples from SHIDC-B-Ki-67. Red points shows negative Ki-67 stained tumor cells. Blue points depicts positive Ki-67 stained tumor cells and dots with cyan color, shows tumor infiltrating lymphocytes.  }
\label{fig:qual_res_2}
\end{figure}

\textbf{Qualitative Results. } We further visualize the qualitative results on the SHIDC-BC-Ki-67 test set in Fig.\ref{fig:qual_res_2}. We compare the detection and classification results of our proposed pipeline with different backends besides the ground truth and the input image. The proposed pipeline leads to the detection of highly overlapped cells with different colors, sizes, and lighting conditions as diverse samples shown in Fig.\ref{fig:qual_res_2} .

\textbf{Limitations.} Besides the fact that tumoral cell size and color may vary from patient to patient, using only Ki-67 marked images for TIL scoring makes this task even harder. To the TIL scoring end, experts only annotated the cells as TIL where they seems similar. This approach may be limited in this regard, but thanks to the powerful deep model feature extractors, we believe that we can have an estimation of TIL score based on Ki-67 marked samples instead of using dual stained or TIL marked images. 

\textbf{Code Availability.}
Codes and scripts to replicate the results is provided on the related repository \url{https://github.com/SHIDCenter/PathoNet}. 

\section*{Conclusion}
In this article we introduced a new benchmark for cell detection, classification , proliferation index estimation, and TIL scoring using histopathological images. We further proposed a new pipeline to be used in cell detection and classification that can utilize different deep models for feature extraction. We evaluated this pipeline on immunonegative, immunopositive, and TIL cell detection on Ki-67 stained images. Also, we suggested a residual inception and designed a new light-weight backbone called PathoNet that achieved state of the art results on the proposed dataset. The suggested pipeline showed high accuracy in Ki-67 and TIL cut-off classification with all backends. Finally, we showed that, designing PathoNet using RDIM provides high accuracy while slightly increases model parameters. 

\section*{Data Availability}
All analysis results and ablations withing this study are included is this article. In addition, dataset is publicly available and can be accessed from \url{http://www.SHIDC.ir}.

\newpage


\newpage
\section*{Acknowledgements}
We would like to acknowledge Maryam Amin Safae for her efforts. 

\section*{Author Information}
\subsection*{Affiliations}

\textbf{Trauma Research Center, Shiraz University of Medical Sciences, Shiraz, Iran} \newline
\textbf{Molecular Pathology and Cytogenetics division, Department of Pathology, Shiraz University of Medical Sciences, Shiraz, Iran.} \newline
Amirreza Dehghanian \newline

\noindent\textbf{Department of Pathology, Shiraz University of Medical Science, Shiraz, Iran} \newline Bita Pakniyat Jahromi and Mahsa Kohandel Shirazi \newline

\noindent\textbf{Shiraz University of Medical Science, Shiraz, Iran}\newline Fateme Movahedi and Shayan Majidi \newline

\noindent\textbf{Department of Clinical Pharmacy, Shiraz University of Medical Sciences, Shiraz, Iran} \newline Dena Firouzabadi \newline

\noindent\textbf{Department of Computer Science and Engineering, Shiraz University, Shiraz, Iran.}\newline Rasool Sabzi \newline

\noindent\textbf{Department of Computer Science and Engineering, Shiraz University, Shiraz, Iran. (During this study)}\newline
\noindent\textbf{Department of Computer Science and Engineering, Koc University, Istanbul, Turkey. (Current)}\newline Farzin Negahbani  \newline

\subsection*{Contributions}
B.P. Jahromi and A. Dehghanian prepared the slides and determined hotspots. F. Negahbani and R. Sabzi performed machine learning experiments plus slide imaging. A. Dehghanian and B.P. Jahromi supervised and verified image gathering. B.P. Jahromi, M.K. Shirazi, F. Movahedi, and S. Majidi carried out labeling and A. Dehghanian and D, Firouzabadi further validated labels. A. Dehghanian, F. Negahbani, R. Sabzi, and D. Firouzabadi wrote the manuscript and all authors reviewed the manuscript.

\subsection*{Corresponding Authors}
Correspondence to Amirreza Dehghanian.

\section*{Additional information}

\subsection*{Competing interests}
The authors declare no competing interests.

\newpage
\appendix
\titleformat{\section}[block]{\Large\bfseries}{\appendixname~\thesection}{1em}{}

\section{}\label{app:A}

This appendix provides a supplementary document containing detailed information about data preprocessing, parameters, threshold tuning, web-based tool (SHIDC-Lab) for labeling, and imaging setup information.

\subsection{Imaging Setup}
This camera uses an image sensor: Aptina 1/2.3 inch color CMOS with a resolution equal to 4912x3684 pixels (18 MP) with pixel size about 1.25um x 1.25um; Complementary details for the mentioned camera are as follows: SNR:36.3 dB; Dynamic Range: 65.8 dB; Frame speed: 5.6 FPS at 4912x3684, 18.1 FPS at 2456x1842, 32.2 FPS at 1228x922.

\subsection{SHIDC-Lab Software}
SHIDC-Lab is designed to provide flexibility, speed, easy monitoring capability, and revision ability that can be run on any device with a web browser so that experts can perform labeling in their spare time. This framework is written in a general manner, which can provide annotation tools for other datasets as well. In SHIDC-Lab software, the process starts by uploading images to the server and defining annotation classes, corresponding signs, and colors.  Afterward, the head expert assigns images to each expert to perform labeling. After that, experts, regarding their defined authorizations, perform image labeling and revision of already annotated images.

\subsection{Data Label Representation}
\begin{figure}[thpb]
\centering
\includegraphics[width=0.4\linewidth]{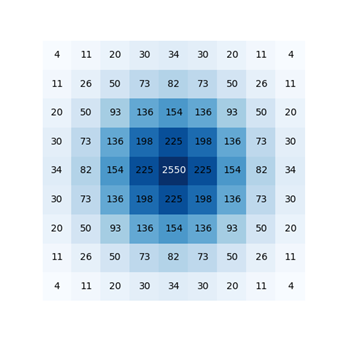}
\caption{Gaussian distribution label visualization. Center value has higher score to stimulate the network to choose center pixel rather than an adjacent pixel.}
\label{fig:label_dist}
\end{figure}

After determining a center pixel by the experts, an abnormal Gaussian distribution fits on the center with a variance of 9 pixels and the maximum value equal to 255. Further, we observed that we could motivate the network to find the center pixels by increasing the center pixel's maximum value. The maximum value in this study was 2250. Fig. \ref{fig:label_dist} depicts the Gaussian distribution label and pixel values. 

\subsection{Preprocessing and training}
In order to prepare data, we first crop the image and its corresponding label to 256 x 256 pixels. Then we split the data into two parts; 0.7 of the data for training and 0.3 for the test. Further, data augmentation is performed by flipping images with respect to x and y coordinates as well as 90, 180, and 270 degrees rotations. For training the PathoNet, the simple MSE loss function by means of the ADAM optimizer is used. The learning rate is set to 0.0001 and decreases with a 0.1 rate every ten epochs. In this study, the Keras framework was used to train the network using two NVIDIA Geforce GTX 1060 and an Intel Core-i5 6400 processor.

\subsection{Threshold Tuning}
As explained in the Methodology section, to extract cell center pixels from the network's density map, thresholding is applied. To tune this threshold for each model, all values from 0 to 255 with a step size of 5 have been evaluated in terms of the cell predictions' F1 score and the best threshold picked for each model. Table. \ref{tab:thresholds} shows the tuned thresholds in this study. 

\begin{table}[h]
\centering
\begin{tabular}{|c|c|c|c|}
\hline
\multirow{2}{*}{\textbf{Model}} & \multicolumn{3}{c|}{\textbf{Cell Type}} \\ \cline{2-4} 
 & \multicolumn{1}{l|}{Positive} & \multicolumn{1}{l|}{Negative} & \multicolumn{1}{l|}{TIL} \\ \hline
Modified DeepLabv3-Mobilenetv2 & 45 & 35 & 5 \\ \hline
Modified DeepLabv3-Xeption & 95 & 80 & 75 \\ \hline
Modified FCRN-A & 120 & 125 & 65 \\ \hline
Modified FCRN-B & 120 & 125 & 70 \\ \hline
Ours (PathoNet) & 120 & 180 & 40 \\ \hline
\end{tabular}
\caption{Threshold values used in this study.}
\label{tab:thresholds}
\end{table}

\subsection{Watershed Algorithm Parameters}
In this study, the implementation of the watershed algorithm from the Skimage library was used. To improve the watershed algorithm's accuracy, maximum points of the density map passed to the algorithm as the initial minimum points. As another parameter to this algorithm, the minimum distance between two maximum points is picked as 5.


\begin{thebibliography}{10}
\urlstyle{rm}
\expandafter\ifx\csname url\endcsname\relax
  \def\url#1{\texttt{#1}}\fi
\expandafter\ifx\csname urlprefix\endcsname\relax\def\urlprefix{URL }\fi
\expandafter\ifx\csname doiprefix\endcsname\relax\def\doiprefix{DOI: }\fi
\providecommand{\bibinfo}[2]{#2}
\providecommand{\eprint}[2][]{\url{#2}}

\bibitem{gerdes1983production}
\bibinfo{author}{Gerdes, J.}, \bibinfo{author}{Schwab, U.},
  \bibinfo{author}{Lemke, H.} \& \bibinfo{author}{Stein, H.}
\newblock \bibinfo{journal}{\bibinfo{title}{Production of a mouse monoclonal
  antibody reactive with a human nuclear antigen associated with cell
  proliferation}}.
\newblock {\emph{\JournalTitle{International journal of cancer}}}
  \textbf{\bibinfo{volume}{31}}, \bibinfo{pages}{13--20}
  (\bibinfo{year}{1983}).

\bibitem{gerdes1984cell}
\bibinfo{author}{Gerdes, J.} \emph{et~al.}
\newblock \bibinfo{journal}{\bibinfo{title}{Cell cycle analysis of a cell
  proliferation-associated human nuclear antigen defined by the monoclonal
  antibody ki-67.}}
\newblock {\emph{\JournalTitle{The journal of immunology}}}
  \textbf{\bibinfo{volume}{133}}, \bibinfo{pages}{1710--1715}
  (\bibinfo{year}{1984}).

\bibitem{lopez1991modalities}
\bibinfo{author}{Lopez, F.} \emph{et~al.}
\newblock \bibinfo{journal}{\bibinfo{title}{Modalities of synthesis of ki67
  antigen during the stimulation of lymphocytes}}.
\newblock {\emph{\JournalTitle{Cytometry: The Journal of the International
  Society for Analytical Cytology}}} \textbf{\bibinfo{volume}{12}},
  \bibinfo{pages}{42--49} (\bibinfo{year}{1991}).

\bibitem{siegel2015cancer}
\bibinfo{author}{Siegel, R.~L.}, \bibinfo{author}{Miller, K.~D.} \&
  \bibinfo{author}{Jemal, A.}
\newblock \bibinfo{journal}{\bibinfo{title}{Cancer statistics, 2015}}.
\newblock {\emph{\JournalTitle{CA: a cancer journal for clinicians}}}
  \textbf{\bibinfo{volume}{65}}, \bibinfo{pages}{5--29} (\bibinfo{year}{2015}).

\bibitem{dowsett2008emerging}
\bibinfo{author}{Dowsett, M.} \& \bibinfo{author}{Dunbier, A.~K.}
\newblock \bibinfo{journal}{\bibinfo{title}{Emerging biomarkers and new
  understanding of traditional markers in personalized therapy for breast
  cancer}}.
\newblock {\emph{\JournalTitle{Clinical Cancer Research}}}
  \textbf{\bibinfo{volume}{14}}, \bibinfo{pages}{8019--8026}
  (\bibinfo{year}{2008}).

\bibitem{jones2009prognostic}
\bibinfo{author}{Jones, R.~L.} \emph{et~al.}
\newblock \bibinfo{journal}{\bibinfo{title}{The prognostic significance of ki67
  before and after neoadjuvant chemotherapy in breast cancer}}.
\newblock {\emph{\JournalTitle{Breast cancer research and treatment}}}
  \textbf{\bibinfo{volume}{116}}, \bibinfo{pages}{53--68}
  (\bibinfo{year}{2009}).

\bibitem{taneja2010classical}
\bibinfo{author}{Taneja, P.} \emph{et~al.}
\newblock \bibinfo{journal}{\bibinfo{title}{Classical and novel prognostic
  markers for breast cancer and their clinical significance}}.
\newblock {\emph{\JournalTitle{Clinical Medicine Insights: Oncology}}}
  \textbf{\bibinfo{volume}{4}}, \bibinfo{pages}{CMO--S4773}
  (\bibinfo{year}{2010}).

\bibitem{urruticoechea2005proliferation}
\bibinfo{author}{Urruticoechea, A.}, \bibinfo{author}{Smith, I.~E.} \&
  \bibinfo{author}{Dowsett, M.}
\newblock \bibinfo{journal}{\bibinfo{title}{Proliferation marker ki-67 in early
  breast cancer}}.
\newblock {\emph{\JournalTitle{Journal of clinical oncology}}}
  \textbf{\bibinfo{volume}{23}}, \bibinfo{pages}{7212--7220}
  (\bibinfo{year}{2005}).

\bibitem{dowsett2011assessment}
\bibinfo{author}{Dowsett, M.} \emph{et~al.}
\newblock \bibinfo{journal}{\bibinfo{title}{Assessment of ki67 in breast
  cancer: recommendations from the international ki67 in breast cancer working
  group}}.
\newblock {\emph{\JournalTitle{Journal of the National Cancer Institute}}}
  \textbf{\bibinfo{volume}{103}}, \bibinfo{pages}{1656--1664}
  (\bibinfo{year}{2011}).

\bibitem{denkert2010tumor}
\bibinfo{author}{Denkert, C.} \emph{et~al.}
\newblock \bibinfo{journal}{\bibinfo{title}{Tumor-associated lymphocytes as an
  independent predictor of response to neoadjuvant chemotherapy in breast
  cancer}}.
\newblock {\emph{\JournalTitle{J Clin Oncol}}} \textbf{\bibinfo{volume}{28}},
  \bibinfo{pages}{105--113} (\bibinfo{year}{2010}).

\bibitem{denkert2018tumour}
\bibinfo{author}{Denkert, C.} \emph{et~al.}
\newblock \bibinfo{journal}{\bibinfo{title}{Tumour-infiltrating lymphocytes and
  prognosis in different subtypes of breast cancer: a pooled analysis of 3771
  patients treated with neoadjuvant therapy}}.
\newblock {\emph{\JournalTitle{The lancet oncology}}}
  \textbf{\bibinfo{volume}{19}}, \bibinfo{pages}{40--50}
  (\bibinfo{year}{2018}).

\bibitem{mao2014value}
\bibinfo{author}{Mao, Y.} \emph{et~al.}
\newblock \bibinfo{journal}{\bibinfo{title}{The value of tumor infiltrating
  lymphocytes (tils) for predicting response to neoadjuvant chemotherapy in
  breast cancer: a systematic review and meta-analysis}}.
\newblock {\emph{\JournalTitle{PloS one}}} \textbf{\bibinfo{volume}{9}},
  \bibinfo{pages}{e115103} (\bibinfo{year}{2014}).

\bibitem{mao2016prognostic}
\bibinfo{author}{Mao, Y.} \emph{et~al.}
\newblock \bibinfo{journal}{\bibinfo{title}{The prognostic value of
  tumor-infiltrating lymphocytes in breast cancer: a systematic review and
  meta-analysis}}.
\newblock {\emph{\JournalTitle{PloS one}}} \textbf{\bibinfo{volume}{11}},
  \bibinfo{pages}{e0152500} (\bibinfo{year}{2016}).

\bibitem{kononenko1997application}
\bibinfo{author}{Kononenko, I.}, \bibinfo{author}{Bratko, I.} \&
  \bibinfo{author}{Kukar, M.}
\newblock \bibinfo{journal}{\bibinfo{title}{Application of machine learning to
  medical diagnosis}}.
\newblock {\emph{\JournalTitle{Machine Learning and Data Mining: Methods and
  Applications}}} \textbf{\bibinfo{volume}{389}}, \bibinfo{pages}{408}
  (\bibinfo{year}{1997}).

\bibitem{soanssa}
\bibinfo{author}{Soans, N.}, \bibinfo{author}{Asali, E.},
  \bibinfo{author}{Hong, Y.} \& \bibinfo{author}{Doshi, P.}
\newblock \bibinfo{title}{Sa-net: Robust state-action recognition for learning
  from observations}.
\newblock In \emph{\bibinfo{booktitle}{IEEE International Conference on
  Robotics and Automation (ICRA)}}, \bibinfo{pages}{2153--2159}
  (\bibinfo{year}{2020}).

\bibitem{haskins2020deep}
\bibinfo{author}{Haskins, G.}, \bibinfo{author}{Kruger, U.} \&
  \bibinfo{author}{Yan, P.}
\newblock \bibinfo{journal}{\bibinfo{title}{Deep learning in medical image
  registration: A survey}}.
\newblock {\emph{\JournalTitle{Machine Vision and Applications}}}
  \textbf{\bibinfo{volume}{31}}, \bibinfo{pages}{8} (\bibinfo{year}{2020}).

\bibitem{hafiz2020survey}
\bibinfo{author}{Hafiz, A.~M.} \& \bibinfo{author}{Bhat, G.~M.}
\newblock \bibinfo{title}{A survey of deep learning techniques for medical
  diagnosis}.
\newblock In \emph{\bibinfo{booktitle}{Information and Communication Technology
  for Sustainable Development}}, \bibinfo{pages}{161--170}
  (\bibinfo{publisher}{Springer}, \bibinfo{year}{2020}).

\bibitem{doi:10.1162/neco.1989.1.4.541}
\bibinfo{author}{LeCun, Y.} \emph{et~al.}
\newblock \bibinfo{journal}{\bibinfo{title}{Backpropagation applied to
  handwritten zip code recognition}}.
\newblock {\emph{\JournalTitle{Neural Computation}}}
  \textbf{\bibinfo{volume}{1}}, \bibinfo{pages}{541--551},
  \doiprefix\url{10.1162/neco.1989.1.4.541} (\bibinfo{year}{1989}).
\newblock \eprint{https://doi.org/10.1162/neco.1989.1.4.541}.

\bibitem{ILSVRC15}
\bibinfo{author}{Russakovsky, O.} \emph{et~al.}
\newblock \bibinfo{journal}{\bibinfo{title}{{ImageNet Large Scale Visual
  Recognition Challenge}}}.
\newblock {\emph{\JournalTitle{International Journal of Computer Vision
  (IJCV)}}} \textbf{\bibinfo{volume}{115}}, \bibinfo{pages}{211--252},
  \doiprefix\url{10.1007/s11263-015-0816-y} (\bibinfo{year}{2015}).

\bibitem{krizhevsky2012imagenet}
\bibinfo{author}{Krizhevsky, A.}, \bibinfo{author}{Sutskever, I.} \&
  \bibinfo{author}{Hinton, G.~E.}
\newblock \bibinfo{title}{Imagenet classification with deep convolutional
  neural networks}.
\newblock In \emph{\bibinfo{booktitle}{Advances in neural information
  processing systems}}, \bibinfo{pages}{1097--1105} (\bibinfo{year}{2012}).

\bibitem{xing2013automatic}
\bibinfo{author}{Xing, F.}, \bibinfo{author}{Su, H.}, \bibinfo{author}{Neltner,
  J.} \& \bibinfo{author}{Yang, L.}
\newblock \bibinfo{journal}{\bibinfo{title}{Automatic ki-67 counting using
  robust cell detection and online dictionary learning}}.
\newblock {\emph{\JournalTitle{IEEE Transactions on Biomedical Engineering}}}
  \textbf{\bibinfo{volume}{61}}, \bibinfo{pages}{859--870}
  (\bibinfo{year}{2013}).

\bibitem{swiderska2015hot}
\bibinfo{author}{Swiderska, Z.}, \bibinfo{author}{Markiewicz, T.},
  \bibinfo{author}{Grala, B.} \& \bibinfo{author}{Slodkowska, J.}
\newblock \bibinfo{title}{Hot-spot selection and evaluation methods for whole
  slice images of meningiomas and oligodendrogliomas}.
\newblock In \emph{\bibinfo{booktitle}{2015 37th Annual International
  Conference of the IEEE Engineering in Medicine and Biology Society (EMBC)}},
  \bibinfo{pages}{6252--6256} (\bibinfo{organization}{IEEE},
  \bibinfo{year}{2015}).

\bibitem{shi2016automated}
\bibinfo{author}{Shi, P.} \emph{et~al.}
\newblock \bibinfo{journal}{\bibinfo{title}{Automated ki-67 quantification of
  immunohistochemical staining image of human nasopharyngeal carcinoma
  xenografts}}.
\newblock {\emph{\JournalTitle{Scientific reports}}}
  \textbf{\bibinfo{volume}{6}}, \bibinfo{pages}{32127} (\bibinfo{year}{2016}).

\bibitem{geread2019ihc}
\bibinfo{author}{Geread, R.~S.} \emph{et~al.}
\newblock \bibinfo{journal}{\bibinfo{title}{Ihc colour histograms for
  unsupervised ki67 proliferation index calculation}}.
\newblock {\emph{\JournalTitle{Frontiers in bioengineering and biotechnology}}}
  \textbf{\bibinfo{volume}{7}}, \bibinfo{pages}{226} (\bibinfo{year}{2019}).

\bibitem{xu2014deep}
\bibinfo{author}{Xu, Y.} \emph{et~al.}
\newblock \bibinfo{title}{Deep learning of feature representation with multiple
  instance learning for medical image analysis}.
\newblock In \emph{\bibinfo{booktitle}{2014 IEEE international conference on
  acoustics, speech and signal processing (ICASSP)}},
  \bibinfo{pages}{1626--1630} (\bibinfo{organization}{IEEE},
  \bibinfo{year}{2014}).

\bibitem{weidi2015microscopy}
\bibinfo{author}{Weidi, X.}, \bibinfo{author}{Noble, J.~A.} \&
  \bibinfo{author}{Zisserman, A.}
\newblock \bibinfo{title}{Microscopy cell counting with fully convolutional
  regression networks}.
\newblock In \emph{\bibinfo{booktitle}{1st Deep Learning Workshop, Medical
  Image Computing and Computer-Assisted Intervention (MICCAI)}}
  (\bibinfo{year}{2015}).

\bibitem{paul2017count}
\bibinfo{author}{Paul~Cohen, J.}, \bibinfo{author}{Boucher, G.},
  \bibinfo{author}{Glastonbury, C.~A.}, \bibinfo{author}{Lo, H.~Z.} \&
  \bibinfo{author}{Bengio, Y.}
\newblock \bibinfo{title}{Count-ception: Counting by fully convolutional
  redundant counting}.
\newblock In \emph{\bibinfo{booktitle}{Proceedings of the IEEE International
  Conference on Computer Vision}}, \bibinfo{pages}{18--26}
  (\bibinfo{year}{2017}).

\bibitem{spanhol2017deep}
\bibinfo{author}{Spanhol, F.~A.}, \bibinfo{author}{Oliveira, L.~S.},
  \bibinfo{author}{Cavalin, P.~R.}, \bibinfo{author}{Petitjean, C.} \&
  \bibinfo{author}{Heutte, L.}
\newblock \bibinfo{title}{Deep features for breast cancer histopathological
  image classification}.
\newblock In \emph{\bibinfo{booktitle}{2017 IEEE International Conference on
  Systems, Man, and Cybernetics (SMC)}}, \bibinfo{pages}{1868--1873}
  (\bibinfo{organization}{IEEE}, \bibinfo{year}{2017}).

\bibitem{saha2017jiang}
\bibinfo{author}{Saha, M.}, \bibinfo{author}{Chakraborty, C.},
  \bibinfo{author}{Arun, I.}, \bibinfo{author}{Ahmed, R.} \&
  \bibinfo{author}{Chatterjee, S.}
\newblock \bibinfo{journal}{\bibinfo{title}{An advanced deep learning approach
  for ki-67 stained hotspot detection and proliferation rate scoring for
  prognostic evaluation of breast cancer}}.
\newblock {\emph{\JournalTitle{Scientific reports}}}
  \textbf{\bibinfo{volume}{7}}, \bibinfo{pages}{3213} (\bibinfo{year}{2017}).

\bibitem{zhang2018tumor}
\bibinfo{author}{Zhang, R.} \emph{et~al.}
\newblock \bibinfo{journal}{\bibinfo{title}{Tumor cell identification in ki-67
  images on deep learning}}.
\newblock {\emph{\JournalTitle{Molecular \& Cellular Biomechanics}}}
  \textbf{\bibinfo{volume}{15}}, \bibinfo{pages}{177--187}
  (\bibinfo{year}{2018}).

\bibitem{sornapudi2018deep}
\bibinfo{author}{Sornapudi, S.} \emph{et~al.}
\newblock \bibinfo{journal}{\bibinfo{title}{Deep learning nuclei detection in
  digitized histology images by superpixels}}.
\newblock {\emph{\JournalTitle{Journal of pathology informatics}}}
  \textbf{\bibinfo{volume}{9}} (\bibinfo{year}{2018}).

\bibitem{jiang2019breast}
\bibinfo{author}{Jiang, Y.}, \bibinfo{author}{Chen, L.},
  \bibinfo{author}{Zhang, H.} \& \bibinfo{author}{Xiao, X.}
\newblock \bibinfo{journal}{\bibinfo{title}{Breast cancer histopathological
  image classification using convolutional neural networks with small se-resnet
  module}}.
\newblock {\emph{\JournalTitle{PloS one}}} \textbf{\bibinfo{volume}{14}},
  \bibinfo{pages}{e0214587} (\bibinfo{year}{2019}).

\bibitem{liu2019novel}
\bibinfo{author}{Liu, Q.}, \bibinfo{author}{Junker, A.},
  \bibinfo{author}{Murakami, K.} \& \bibinfo{author}{Hu, P.}
\newblock \bibinfo{title}{A novel convolutional regression network for cell
  counting}.
\newblock In \emph{\bibinfo{booktitle}{2019 IEEE 7th International Conference
  on Bioinformatics and Computational Biology (ICBCB)}},
  \bibinfo{pages}{44--49} (\bibinfo{organization}{IEEE}, \bibinfo{year}{2019}).

\bibitem{spanhol2016breast}
\bibinfo{author}{Spanhol, F.~A.}, \bibinfo{author}{Oliveira, L.~S.},
  \bibinfo{author}{Petitjean, C.} \& \bibinfo{author}{Heutte, L.}
\newblock \bibinfo{title}{Breast cancer histopathological image classification
  using convolutional neural networks}.
\newblock In \emph{\bibinfo{booktitle}{2016 international joint conference on
  neural networks (IJCNN)}}, \bibinfo{pages}{2560--2567}
  (\bibinfo{organization}{IEEE}, \bibinfo{year}{2016}).

\bibitem{lempitsky2010learning}
\bibinfo{author}{Lempitsky, V.} \& \bibinfo{author}{Zisserman, A.}
\newblock \bibinfo{title}{Learning to count objects in images}.
\newblock In \emph{\bibinfo{booktitle}{Advances in neural information
  processing systems}}, \bibinfo{pages}{1324--1332} (\bibinfo{year}{2010}).

\bibitem{kainz2015you}
\bibinfo{author}{Kainz, P.}, \bibinfo{author}{Urschler, M.},
  \bibinfo{author}{Schulter, S.}, \bibinfo{author}{Wohlhart, P.} \&
  \bibinfo{author}{Lepetit, V.}
\newblock \bibinfo{title}{You should use regression to detect cells}.
\newblock In \emph{\bibinfo{booktitle}{International Conference on Medical
  Image Computing and Computer-Assisted Intervention}},
  \bibinfo{pages}{276--283} (\bibinfo{organization}{Springer},
  \bibinfo{year}{2015}).

\bibitem{marsden2018people}
\bibinfo{author}{Marsden, M.}, \bibinfo{author}{McGuinness, K.},
  \bibinfo{author}{Little, S.}, \bibinfo{author}{Keogh, C.~E.} \&
  \bibinfo{author}{O'Connor, N.~E.}
\newblock \bibinfo{title}{People, penguins and petri dishes: Adapting object
  counting models to new visual domains and object types without forgetting}.
\newblock In \emph{\bibinfo{booktitle}{Proceedings of the IEEE Conference on
  Computer Vision and Pattern Recognition}}, \bibinfo{pages}{8070--8079}
  (\bibinfo{year}{2018}).

\bibitem{ronneberger2015u}
\bibinfo{author}{Ronneberger, O.}, \bibinfo{author}{Fischer, P.} \&
  \bibinfo{author}{Brox, T.}
\newblock \bibinfo{title}{U-net: Convolutional networks for biomedical image
  segmentation}.
\newblock In \emph{\bibinfo{booktitle}{International Conference on Medical
  image computing and computer-assisted intervention}},
  \bibinfo{pages}{234--241} (\bibinfo{organization}{Springer},
  \bibinfo{year}{2015}).

\bibitem{myronenko20183d}
\bibinfo{author}{Myronenko, A.}
\newblock \bibinfo{title}{3d mri brain tumor segmentation using autoencoder
  regularization}.
\newblock In \emph{\bibinfo{booktitle}{International MICCAI Brainlesion
  Workshop}}, \bibinfo{pages}{311--320} (\bibinfo{organization}{Springer},
  \bibinfo{year}{2018}).

\bibitem{dolz2018ivd}
\bibinfo{author}{Dolz, J.}, \bibinfo{author}{Desrosiers, C.} \&
  \bibinfo{author}{Ayed, I.~B.}
\newblock \bibinfo{title}{Ivd-net: Intervertebral disc localization and
  segmentation in mri with a multi-modal unet}.
\newblock In \emph{\bibinfo{booktitle}{International Workshop and Challenge on
  Computational Methods and Clinical Applications for Spine Imaging}},
  \bibinfo{pages}{130--143} (\bibinfo{organization}{Springer},
  \bibinfo{year}{2018}).

\bibitem{szegedy2015going}
\bibinfo{author}{Szegedy, C.} \emph{et~al.}
\newblock \bibinfo{title}{Going deeper with convolutions}.
\newblock In \emph{\bibinfo{booktitle}{Proceedings of the IEEE conference on
  computer vision and pattern recognition}}, \bibinfo{pages}{1--9}
  (\bibinfo{year}{2015}).

\bibitem{yang2019dilated}
\bibinfo{author}{Yang, S.}, \bibinfo{author}{Lin, G.}, \bibinfo{author}{Jiang,
  Q.} \& \bibinfo{author}{Lin, W.}
\newblock \bibinfo{title}{A dilated inception network for visual saliency
  prediction} (\bibinfo{year}{2019}).
\newblock \eprint{1904.03571}.

\bibitem{he2016deep}
\bibinfo{author}{He, K.}, \bibinfo{author}{Zhang, X.}, \bibinfo{author}{Ren,
  S.} \& \bibinfo{author}{Sun, J.}
\newblock \bibinfo{title}{Deep residual learning for image recognition}.
\newblock In \emph{\bibinfo{booktitle}{Proceedings of the IEEE conference on
  computer vision and pattern recognition}}, \bibinfo{pages}{770--778}
  (\bibinfo{year}{2016}).

\bibitem{atta20163d}
\bibinfo{author}{Atta-Fosu, T.} \emph{et~al.}
\newblock \bibinfo{journal}{\bibinfo{title}{3d clumped cell segmentation using
  curvature based seeded watershed}}.
\newblock {\emph{\JournalTitle{Journal of imaging}}}
  \textbf{\bibinfo{volume}{2}}, \bibinfo{pages}{31} (\bibinfo{year}{2016}).

\bibitem{Lantujoul1978LaSE}
\bibinfo{author}{Lantu{\'e}joul, C.}
\newblock \bibinfo{title}{La squelettisation et son application aux mesures
  topologiques des mosaiques polycristallines} (\bibinfo{year}{1978}).

\bibitem{kornilov2018overview}
\bibinfo{author}{Kornilov, A.~S.} \& \bibinfo{author}{Safonov, I.~V.}
\newblock \bibinfo{journal}{\bibinfo{title}{An overview of watershed algorithm
  implementations in open source libraries}}.
\newblock {\emph{\JournalTitle{Journal of Imaging}}}
  \textbf{\bibinfo{volume}{4}}, \bibinfo{pages}{123} (\bibinfo{year}{2018}).

\bibitem{chen2017rethinking}
\bibinfo{author}{Chen, L.-C.}, \bibinfo{author}{Papandreou, G.},
  \bibinfo{author}{Schroff, F.} \& \bibinfo{author}{Adam, H.}
\newblock \bibinfo{journal}{\bibinfo{title}{Rethinking atrous convolution for
  semantic image segmentation}}.
\newblock {\emph{\JournalTitle{arXiv preprint arXiv:1706.05587}}}
  (\bibinfo{year}{2017}).

\bibitem{Everingham10}
\bibinfo{author}{Everingham, M.}, \bibinfo{author}{Van~Gool, L.},
  \bibinfo{author}{Williams, C. K.~I.}, \bibinfo{author}{Winn, J.} \&
  \bibinfo{author}{Zisserman, A.}
\newblock \bibinfo{journal}{\bibinfo{title}{The pascal visual object classes
  (voc) challenge}}.
\newblock {\emph{\JournalTitle{International Journal of Computer Vision}}}
  \textbf{\bibinfo{volume}{88}}, \bibinfo{pages}{303--338}
  (\bibinfo{year}{2010}).

\bibitem{doi:10.1080/21681163.2016.1149104}
\bibinfo{author}{Xie, W.}, \bibinfo{author}{Noble, J.~A.} \&
  \bibinfo{author}{Zisserman, A.}
\newblock \bibinfo{journal}{\bibinfo{title}{Microscopy cell counting and
  detection with fully convolutional regression networks}}.
\newblock {\emph{\JournalTitle{Computer Methods in Biomechanics and Biomedical
  Engineering: Imaging \& Visualization}}} \textbf{\bibinfo{volume}{6}},
  \bibinfo{pages}{283--292}, \doiprefix\url{10.1080/21681163.2016.1149104}
  (\bibinfo{year}{2018}).
\newblock \eprint{https://doi.org/10.1080/21681163.2016.1149104}.

\bibitem{chollet2015keras}
\bibinfo{author}{Chollet, F.} \emph{et~al.}
\newblock \bibinfo{title}{Keras}.
\newblock \bibinfo{howpublished}{\url{https://keras.io}}
  (\bibinfo{year}{2015}).

\end{thebibliography}
\end{document}